\documentclass[conference]{IEEEtran}
\usepackage[numbers]{natbib}
\usepackage{amsmath,amssymb,amsfonts,bm}
\usepackage{algorithmic}
\usepackage{graphicx}
\usepackage{textcomp}
\usepackage{xcolor}
\usepackage{hyperref}
\hypersetup{hidelinks}
\usepackage{booktabs}
\usepackage{subcaption}
\usepackage{float}
\usepackage{multirow}
\usepackage{dblfloatfix}   

\def\BibTeX{{\rm B\kern-.05em{\sc i\kern-.025em b}\kern-.08em
    T\kern-.1667em\lower.7ex\hbox{E}\kern-.125emX}}

\begin{document}

\title{End-to-End Parametric Portfolio Policies for Cross-Asset Futures Timing:\\ When Do AI Models Beat Simple Rules?
}

\author{
\IEEEauthorblockN{Austin Pollok}
\IEEEauthorblockA{\textit{Data Science and Operations} \\
\textit{Finance and Business Economics} \\
\textit{USC}\\
Los Angeles, USA \\
pollok@usc.edu}
\and
\IEEEauthorblockN{Kevin Robik}
\IEEEauthorblockA{\textit{Managing Partner} \\
\textit{Critical Technologies, LLC}\\
New York, USA \\
krobik@criticaltechnologies.net}}

\maketitle

\begin{abstract}
Timing-based tilts across asset classes can drive much of the risk and return of a diversified cross-asset portfolio. The standard approach forecasts returns and then optimizes weights. We instead study an end-to-end AI-based policy that maps market states directly to portfolio weights, and we then ask when this one-step modeling approach outperforms simple rules-based strategies. We train these policies on the sixteen most liquid CME futures, where an edge is unlikely to be due to illiquidity, using a differentiable Sharpe ratio loss function, and we benchmark them against equal weighting, risk parity, and time-series momentum. The learned policies rank above the rules on the pooled cross-asset portfolio and in several sub-asset classes, but not uniformly. In gross terms, an LSTM and a transformer-based architecture perform comparably out-of-sample, but diverge when we consider transaction costs. The transformer generates the stronger learned policy, trades far less than the LSTM, and matches or exceeds equal weighting through moderate cost. 

\end{abstract}

\section{Introduction}

Systematic multi-asset investing allocates capital across equities, fixed income, currencies, and commodities, and is a core activity of global-macro and managed-futures strategies. In these portfolios, timing decisions can drive much of the realized risk and return, yet timing asset-class returns is notoriously difficult. Many multi-asset allocation programs therefore rely either on discretionary judgment or on simple systematic rules such as equal weighting, risk parity, and time-series momentum. This paper asks a deliberately practical question: when does an AI-based allocation policy outperform these simple rules by enough to justify its added complexity?

The standard quantitative approach to portfolio construction is still rooted in a two-step mean--variance optimization framework~\cite{markowitz1952}. That framework remains foundational, but in cross-asset timing its weaknesses can be costly. It requires expected-return forecasts, which are noisy~\cite{fama1970}, and it runs the risk of compounding small forecasting errors into unstable weights. In response, a large literature has looked for ways to improve the predict-then-optimize pipeline, including better covariance estimation, richer signals, and alternative objectives. We take a different route, rather than forecast returns and estimate covariances to optimize weights, we learn a state-contingent allocation rule directly.

We employ an approach that follows the end-to-end portfolio-optimization literature~\cite{cucuringu2021universalendtoendapproachportfolio,kisiel2023portfoliotransformers} where a neural network maps market states directly to portfolio weights, and the parameters are trained on the downstream objective itself, here a differentiable Sharpe ratio. This collapses prediction and optimization into a single-step policy. The same idea appears in several fields under different names. In financial economics, it is closely related to parametric portfolio policies, where portfolio weights are modeled directly as functions of characteristics~\cite{brandt2009parametricportfolios,zimmerman2025deepparametricportfolios}. In operations research and management science, the closest analogue is decision-focused or task-based end-to-end optimization, where models are trained against the downstream decision objective rather than an intermediate prediction loss~\cite{donti2017taskbased,elmachtoub2022spo,bertsimas2020predictivePrescriptive}. In reinforcement learning and stochastic optimization, the taxonomy in~\cite{powell2022RLandStochasticOptimization} classifies this as a policy function approximation, which maps states to actions rather than estimating a value function.

This work is closest to the end-to-end trading literature, including Deep Momentum Networks~\cite{lim2019deepmomentum}, deep reinforcement learning for futures~\cite{zhang2019deeprl}, deep portfolio optimization over ETFs~\cite{zhang2020deepportfolio}, and differentiable weight layers for large equity universes~\cite{cucuringu2021universalendtoendapproachportfolio}. Within the transformer literature, the closest papers are~\cite{wood2022momentumtransformer}, who study attention-based futures momentum, and~\cite{kisiel2023portfoliotransformers}, who develop a Portfolio Transformer for Sharpe-optimized asset allocation.

Our contribution is a practitioner-oriented evaluation of this end-to-end framework in a highly liquid, cross-asset investable universe. We study the sixteen most liquid CME futures across six asset classes, benchmark the learned policies against equal weighting, risk parity, and time-series momentum, and evaluate performance under walk-forward training with realistic transaction costs. The goal is not architectural novelty, but to evaluate when an end-to-end policy improves on the simple rules a practitioner would actually run in a global-macro or managed-futures setting, where the cross-section is smaller, the instruments are highly liquid, and the problem is cross-asset timing rather than large-universe portfolio construction.

The evidence is mixed in an informative way. The learned policies rank above the simple rules on the pooled cross-asset portfolio and in several single-asset-class sleeves, but not uniformly. The transformer generally delivers the stronger policy, as it attains higher gross risk-adjusted performance in several universes and trades far less than the LSTM, so its performance survives moderate transaction costs. The LSTM is competitive and is the best model in equities, but its higher turnover erodes much of its advantage once costs are implemented. The strongest results are also not always evidence of market-neutral timing skill. In equities, factor decomposition attributes much of the transformer's performance to market exposure rather than residual alpha.

The main lesson is therefore conditional rather than universal. End-to-end AI allocation can add value, but the value is asset-class dependent, benchmark dependent, and cost dependent. It is strongest where simple rules are weaker and where the model can exploit breadth across contracts. It is least compelling where naive diversification or a standard trend rule already captures most of the available return. Added complexity also does not automatically help. In our tests, seed averaging improves stability, but mixture-of-experts structures, larger feature sets, per-class tuning, and cross-model ensembling do not reliably improve out-of-sample performance.

The rest of the paper proceeds as follows. Section~\ref{sec:data} describes the liquid futures universe and the feature set. Section~\ref{sec:methods} specifies the policy framework, the differentiable Sharpe objective, and the transformer and LSTM architectures. Section~\ref{sec:insample} studies in-sample optimization and model diagnostics. Section~\ref{sec:oos} reports out-of-sample performance, transaction-cost sensitivity, and alpha-beta decompositions. Section~\ref{sec:conclusion} concludes.

\section{Data}
\label{sec:data}

We use end-of-day futures contract data from Barchart, including its exchange-sourced continuous-contract series and roll specification, which provide the history and data quality our setup needs while avoiding look-ahead bias. Our universe is sixteen of the most liquid CME futures, chosen to cover all six major asset classes at the highest liquidity in each (Table~\ref{tab:universe}). The liquidity screen strips out illiquidity premia as a likely source of return, leaving a cleaner test of pure cross-asset timing in the instruments most relevant to global-macro and managed-futures strategies.

Continuous contract series splice successive contract months by rolling on volume and open interest, giving the long, gap-free histories our models need. Rolls, features, benchmarks, and standardizations all use only trailing data, so the evaluation is free of look-ahead biases.

\begin{table}[H]
\centering
\caption{Investable universe: sixteen liquid CME futures across six asset classes.}
\label{tab:universe}
\small
\begin{tabular}{@{}ll@{}}
\toprule
Asset class & Contracts \\
\midrule
Equity Index & ES, NQ, RTY \\
Interest Rates & ZT, ZF, ZN, ZB \\
Foreign Exchange & 6E, 6J \\
Energy & CL, NG \\
Metals & GC, HG, SI \\
Agriculturals & ZC, ZW \\
\bottomrule
\end{tabular}
\end{table}

\begin{figure}[H]
\centering
\includegraphics[width=\columnwidth]{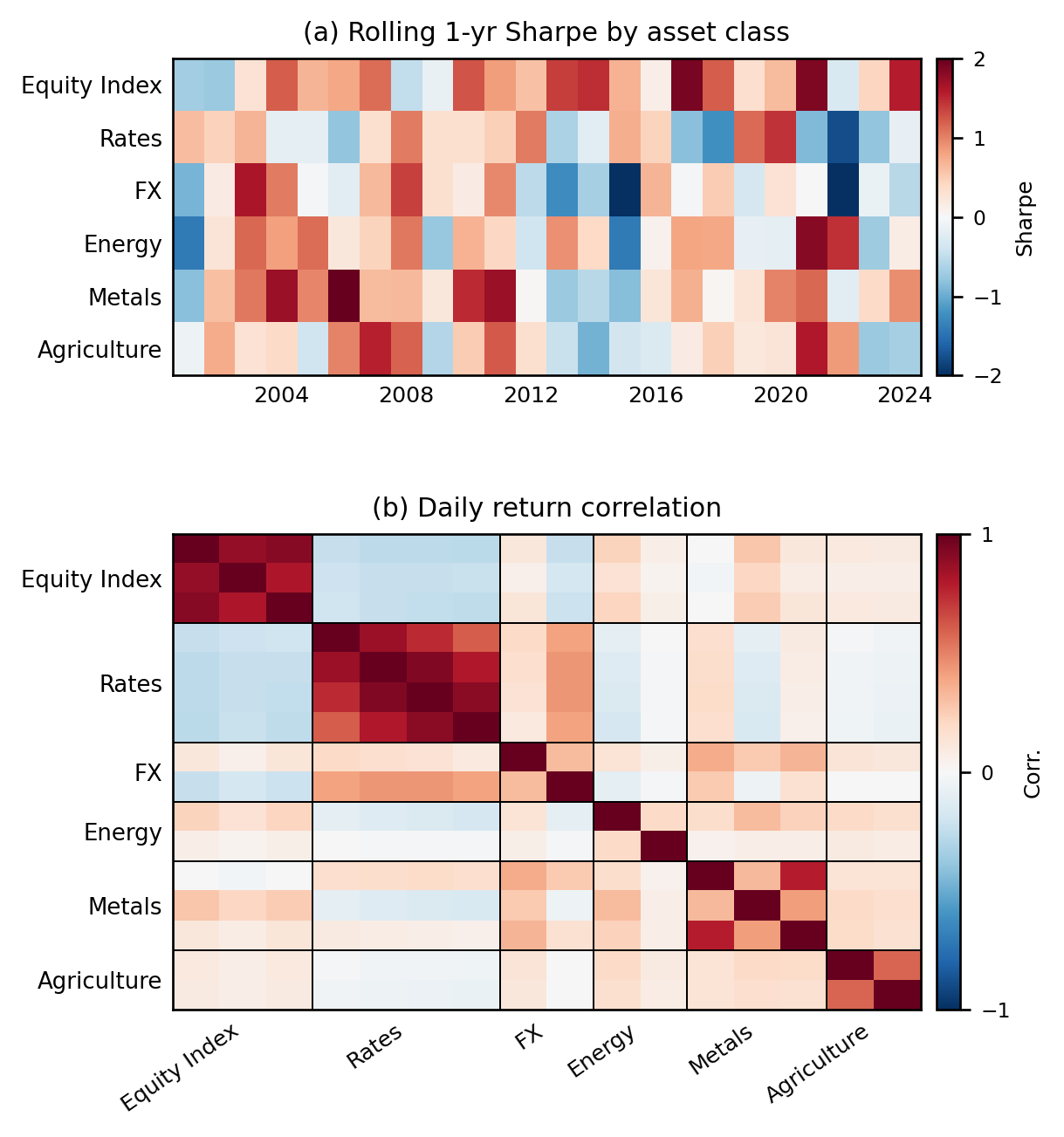}
\caption{Sample period 2001--2024. (a)~Rolling one-year Sharpe ratio of each asset class's equal-weighted portfolio, averaged within each calendar year; every class alternates in sign across years. (b)~Correlation of daily returns among the sixteen contracts, ordered by asset class; the mean pairwise correlation is $0.67$ within an asset class and $0.05$ across, leaving the classes as near-orthogonal return processes.}
\label{fig:hardness}
\end{figure}

Financial returns are hard to model for three important reasons. The signal-to-noise ratio is notably low. The series are nonstationary as means, variances, and correlations drift and break. And the market moves through regimes, trending, mean-reverting, or directionless, under which a learned relationship can reverse. These show up plainly in our cross-asset universe. Within every asset class the rolling one-year Sharpe changes sign repeatedly, and daily returns are highly correlated within an asset class but nearly uncorrelated across asset classes~(Fig.~\ref{fig:hardness}). Any single pooled rule must therefore serve very different, time-varying return processes. 

We construct a deliberately simple set of economically interpretable features intended to capture three dimensions of market behavior: trend, measured by rate of change; mean reversion, measured by lagged autocorrelation; and randomness or regime structure, measured by the Hurst exponent, skewness, and kurtosis. We also include realized volatility and pairwise correlation. Each feature is computed over $1/5/20/60$-day horizons and standardized using rolling $252$-day $z$-scores.

These features play two roles in the paper. First, they provide an in-sample diagnostic for whether the models can learn economically meaningful structure (Section~\ref{sec:insample}). Second, we tested them in the out-of-sample walk-forward setting. In those tests, however, the engineered features added little and transferred inconsistently beyond the raw return history itself. For that reason, most of the reported out-of-sample analysis uses the cross-section of daily returns as the policy state, while a fuller integration of engineered regime features is left to future work.

\section{Models and Methodology}
\label{sec:methods}

We next specify the policy, including the mapping from market states to weights, the differentiable objective used for training, the transformer architecture, and the diagnostics used to monitor optimization.

Following~\cite{cucuringu2021universalendtoendapproachportfolio}, the policy maps market features to per-asset scores and then converts those scores into portfolio weights through a differentiable output layer. The long/short budget constraint is therefore built into the policy itself, rather than imposed as a constraint.

We train the policy by directly maximizing a reward function, defined to be the differentiable Sharpe ratio. Let $R_{i,t}$ denote the realized return of asset $i$ from $t-1$ to $t$, and let $X_t$ denote the vector of market state variables, feature set, observed at time $t$. The portfolio weights are generated by a neural-network-based policy function,
\begin{equation}
\label{eq:policy_function}
w_t(\theta)=\pi_{\theta}(X_t),
\end{equation}
where $\theta$ are the learned policy parameters and $w_t(\theta)=(w_{1,t}(\theta),\ldots,w_{N,t}(\theta))'$. The policy is learned through maximizing the reward function itself, rather than through an intermediate return-forecasting loss. The portfolio return before transaction costs is
\begin{equation}
\label{eq:ret_portfolio}
R_{P,t}(\theta)
=
\sum_{i=1}^{N} w_{i,t-1}(\theta)\, R_{i,t}.
\end{equation}
The training objective is the negative Sharpe ratio,
\begin{equation}
\label{eq:sharpe_obj}
\mathcal{L}_{\text{Sharpe}}(\theta)
=
-
\frac{\mathbb{E}\!\left[R_{P,t}(\theta)\right]}
{\sqrt{\mathbb{V}\!\left[R_{P,t}(\theta)\right]}} .
\end{equation}
This objective is differentiable for gradient-based training because the portfolio weights are differentiable functions of the neural-network parameters~\cite{moody1998,cucuringu2021universalendtoendapproachportfolio}. The headline results optimize this return before transaction costs, so reported gross performance is evaluated before trading costs.

Because the policy can rebalance aggressively from one day to the next, we also study a cost-aware objective that penalizes turnover at coefficient $\lambda$. Returns net of transaction costs are
\begin{equation}
\label{eq:ret_with_costs}
R^{\mathrm{net}}_{P,t}(\theta)
=
R_{P,t}(\theta)
-
\lambda
\sum_{i=1}^{N}
\big|\,w_{i,t-1}(\theta)-w_{i,t-2}(\theta)\,\big| .
\end{equation}
The cost-aware objective minimizes the negative Sharpe ratio of $R^{\mathrm{net}}_{P,t}(\theta)$. Following~\cite{kellymalamudpedersen2024implementableefficientfrontier}, we set $\lambda$ to a realistic $2$ basis points for these liquid contracts and also examine larger values to test the sensitivity of performance to turnover, which also acts as a regularizer. We report gross performance as the main result and net-of-cost performance alongside it in Section~\ref{sec:oos}.

The main policy class is a transformer, following~\cite{kisiel2023portfoliotransformers}. Attention is well suited to long lookbacks, while an LSTM is the other natural sequence model and serves as our neural-network-based baseline. The transformer first embeds the input features, which are daily returns and the regime features described in Section~\ref{sec:data}. Since attention is order-agnostic, we add a Time2Vec embedding, which decomposes the temporal signal into learnable frequencies and phase shifts to capture periodic structure.

The processing block has encoder and decoder components. The encoder applies four identical layers of multi-head self-attention across assets and time, each followed by a gated residual network (GRN), producing a context vector that summarizes the market state. The decoder's four layers attend to this context and produce the output representations used to form portfolio weights. Causal masking ensures that each weight depends only on information available at the time the position is formed. The scaled dot-product attention is
\begin{equation}
\label{eq:attention}
\mathrm{Attention}(Q,K,V)
=
\mathrm{softmax}\!\left(\frac{QK^{\top}}{\sqrt{d_k}} + M\right)V,
\end{equation}
where $Q$, $K$, and $V$ are query, key, and value matrices, $d_k$ is the key dimension, and $M$ is a causal mask with entries equal to $0$ for admissible positions and $-\infty$ for masked positions.

Finally, following~\cite{cucuringu2021universalendtoendapproachportfolio}, the network maps the feature state $X_t$ to decoder scores $s_{i,t}(\theta)$, which are converted into portfolio weights through a signed-softmax layer,
\begin{equation}
\label{eq:post_sigsoftmax}
w_{i,t}(\theta)
=
\mathrm{sign}\!\left(s_{i,t}(\theta)\right)
\frac{\exp\!\left(|s_{i,t}(\theta)|\right)}
{\sum_{j=1}^{N} \exp\!\left(|s_{j,t}(\theta)|\right)} .
\end{equation}
This permits long and short positions while normalizing total absolute exposure across assets.

During training, we monitor loss, gradient norms, and weight norms to ensure stable optimization and to guard against overfitting. We evaluate the resulting policies with standard portfolio metrics, including annualized and rolling Sharpe ratios, turnover, drawdown, and transaction-cost sensitivity.

\section{In-Sample Results: Optimization}
\label{sec:insample}

Before evaluating out-of-sample generalization, we first ask whether the models can learn on the training data alone. This is not an in-sample fit on the full 2000-2024 dataset. It is an in-sample diagnostic inside the walk-forward design, using only the training window available at each point in time. The purpose is to measure the learnability of the data and the capacity of the policy class. This distinction matters for large neural-network policies because poor performance can come from the optimization itself, including unstable gradients, bad local solutions, underfitting, or excessive regularization. If a model cannot improve the training objective on past data, then poor out-of-sample performance may reflect failed optimization. If it fits the training window well but fails out of sample, then failure is more naturally interpreted as weak, unstable, or overfit signal rather than insufficient model capacity.

We run this diagnostic on the initial $40\%$ of the sample, with the remainder of the data being held out for walk-forward validation. We optimize the differentiable negative Sharpe ratio of~\eqref{eq:sharpe_obj} for the two policy classes, a Portfolio Transformer and an LSTM. Because the joint hyperparameter space is too large to grid-search on the transformer, we search the training recipe coordinate-wise on the faster LSTM, which is roughly $30\times$ faster to train, and verify it on the transformer.

Two findings carry into the out-of-sample study. First, the models can learn the training windows and can also overfit them. With an appropriate configuration, including AdamW with a OneCycleLR schedule, GELU activations, pre-activation layer normalization, explicit initialization, and gradient clipping, the in-sample Sharpe improves steadily with a train-validation gap opening once training is pushed too far. This confirms that the policy class has enough capacity to fit the available training data and that early stopping is necessary. Removing normalization or clipping introduces gradient instability.

Second, the engineered features described in Section~\ref{sec:data}, including trend, mean reversion, the Hurst exponent, volatility, correlation, and higher moments, improve in-sample fit modestly, but did not deliver a stable out-of-sample gain in preliminary walk-forward runs. The policy therefore uses the cross-section of daily returns as its state, and we leave the engineered features' full integration to future work. Having tuned mostly on the LSTM, we confirmed rather than re-searched the training hyperparameter configuration on the transformer.

\section{Out-of-Sample Results: Generalization}
\label{sec:oos}

We evaluate the ability of these AI-based policies to generalize out of sample using an expanding-window walk-forward experiment design. Using daily data for the sixteen futures from 2000--2024, each model first trains on the initial $40\%$ of the sample and is then re-fit in roughly four-year test blocks, with a chronological $90/10$ train-validation split and no shuffling. Each block is predicted out of sample, and for the cross-asset portfolio the out-of-sample period runs 2011-2024. We run this once per asset class and once for a single pooled model over all sixteen contracts, the cross-asset universe. We compare the LSTM and transformer policies against three standard rules: equal weight ($1/N$), risk parity, and time-series momentum (TSMOM). All five strategies rebalance daily. Risk parity uses a 60-day inverse-volatility estimate, while TSMOM uses the standard 12-month signal~\cite{lim2019deepmomentum}. Equal weight and risk parity are long-only. The learned policies and TSMOM are long/short with unit gross exposure. Within each universe, all strategies are evaluated over the same out-of-sample window. Because Sharpe ratios are scale-invariant, leverage does not affect the comparison.

\subsection{Performance Across Asset Classes}

Table~\ref{tab:results} reports the out-of-sample performance of every strategy in each universe. We focus first on Sharpe ratios. The transformer has the highest Sharpe in the cross-asset, metals, and agriculture universes. The LSTM leads in the equity index. Equal weighting leads in energy, while time-series momentum leads in interest rates and foreign exchange.

The learned models add the most where the simple rules are weakest. Equal weighting is a well-documented benchmark that is hard to beat~\cite{demiguel2009naive}. It is strongest in long-biased markets, such as the equity indices and the post-2020 energy recovery, where the learned policies do not improve on it. The learned policies perform better in the broad cross-asset universe and in the choppier single-class sleeves.

We also ask which Sharpe ratios are statistically distinguishable from zero. Because we test performance across seven universes, we report Bonferroni-adjusted significance to reduce the chance that one apparently significant result appears simply by testing many universes. Under this correction, the equity-index results for both learned models remain significant, with $t\approx3$. The cross-asset transformer reaches $t\approx2$, which clears the uncorrected $5\%$ level but not the Bonferroni-adjusted threshold.

The economic differences are modest in several places. In the equity index, the learned models barely separate from equal weighting and risk parity, which have Sharpe ratios of $0.78$ and $0.79$. The largest improvement over the simple benchmarks is in agriculture, where the transformer reaches $0.34$ and more than doubles the best long-only strategy. In foreign exchange, the transformer is positive at $0.11$ while the long-only rules are negative, although time-series momentum leads at $0.25$. In the cross-asset universe, the transformer reaches $0.55$, only slightly above equal weighting at $0.52$.

We therefore treat the per-universe rankings as point estimates rather than established differences. The asterisks in Table~\ref{tab:results} test whether each Sharpe ratio differs from zero after the Bonferroni correction. Whether one strategy beats another is a separate comparison. For that question, we use an uncorrected stationary bootstrap on the Sharpe difference. In the cross-asset universe, using a one-month mean block length and $5{,}000$ resamples, the bootstrap assigns probability $0.99$ to the transformer outperforming risk parity and $0.68$ to it outperforming time-series momentum. It assigns only probability $0.59$ to the transformer outperforming equal weighting. The cross-asset transformer therefore matches, rather than significantly beats, $1/N$.

\begin{table*}[t]
\centering
\caption{Out-of-sample risk and return by universe for the two end-to-end models and the three standard rules. All strategies are evaluated on the same walk-forward window within each universe. The benchmark Sharpe ratios match an independent implementation to within rounding. All figures are annualized. \emph{Return} is mean return and \emph{Vol.}\ is volatility, both per unit gross exposure. \emph{Sharpe} is return divided by volatility. \emph{Net Sharpe} is computed after charging $2$~bp on turnover, following the liquid-futures cost assumptions in~\cite{kisiel2023portfoliotransformers,cucuringu2021universalendtoendapproachportfolio}. \emph{Sortino} is return divided by downside deviation. \emph{MDD} is maximum drawdown, and \emph{Calmar} is return divided by MDD. \emph{Turnover} is the mean daily one-sided target-weight change, $\tfrac12\sum_i|w_{i,t}-w_{i,t-1}|$. Constant-weight rules have zero turnover by construction. The best value in each column and panel is bolded. \textsuperscript{*} marks a Sharpe ratio significantly different from zero at the $5\%$ level after a two-sided Bonferroni correction across the seven universes, requiring $|t|>2.69$. The transformer leads on most gross and net metrics. The LSTM is competitive in equities, but its higher turnover erodes its net performance.}
\label{tab:results}
\small
\begin{tabular}{@{}lrrrrrrrr@{}}
\toprule
Strategy & Return & Vol. & Sharpe & Net Sharpe & Sortino & MDD & Calmar & Turnover \\
\midrule
\multicolumn{9}{@{}l}{\textit{Cross-asset}}\\
\midrule
$1/N$ & \textbf{0.05} & 0.09 & 0.52 & 0.52 & 0.73 & 0.24 & 0.20 & \textbf{0.00} \\
Risk parity & 0.01 & \textbf{0.04} & 0.15 & 0.14 & 0.19 & 0.15 & 0.04 & 0.01 \\
TSMOM & 0.03 & 0.08 & 0.37 & 0.35 & 0.53 & 0.21 & 0.14 & 0.02 \\
LSTM & 0.02 & 0.05 & 0.50 & 0.33 & 0.70 & \textbf{0.10} & \textbf{0.26} & 0.17 \\
transformer & 0.05 & 0.08 & \textbf{0.55} & \textbf{0.54} & \textbf{0.74} & 0.27 & 0.17 & 0.02 \\
\midrule
\multicolumn{9}{@{}l}{\textit{Equity index}}\\
\midrule
$1/N$ & 0.15 & 0.19 & 0.78\textsuperscript{*} & 0.78 & 0.98 & 0.40 & 0.37 & \textbf{0.00} \\
Risk parity & \textbf{0.15} & 0.19 & 0.79\textsuperscript{*} & 0.79 & 1.00 & 0.40 & 0.37 & 0.00 \\
TSMOM & 0.07 & 0.16 & 0.45 & 0.44 & 0.56 & 0.32 & 0.23 & 0.02 \\
LSTM & 0.09 & \textbf{0.10} & \textbf{0.87\textsuperscript{*}} & \textbf{0.83} & \textbf{1.02} & \textbf{0.19} & \textbf{0.47} & 0.07 \\
transformer & 0.12 & 0.15 & 0.80\textsuperscript{*} & 0.79 & 0.99 & 0.36 & 0.34 & 0.02 \\
\midrule
\multicolumn{9}{@{}l}{\textit{Interest rates}}\\
\midrule
$1/N$ & -0.00 & 0.05 & -0.09 & -0.09 & -0.14 & 0.27 & -0.02 & \textbf{0.00} \\
Risk parity & -0.01 & 0.03 & -0.19 & -0.19 & -0.26 & 0.16 & -0.03 & 0.00 \\
TSMOM & \textbf{0.02} & 0.05 & \textbf{0.43} & \textbf{0.41} & \textbf{0.62} & 0.10 & \textbf{0.20} & 0.02 \\
LSTM & 0.00 & \textbf{0.02} & 0.09 & -0.23 & 0.11 & \textbf{0.05} & 0.04 & 0.12 \\
transformer & -0.00 & 0.03 & -0.12 & -0.15 & -0.14 & 0.14 & -0.02 & 0.02 \\
\midrule
\multicolumn{9}{@{}l}{\textit{Foreign exchange}}\\
\midrule
$1/N$ & -0.03 & 0.07 & -0.36 & -0.36 & -0.57 & 0.49 & -0.05 & \textbf{0.00} \\
Risk parity & -0.03 & 0.07 & -0.37 & -0.37 & -0.57 & 0.50 & -0.05 & 0.00 \\
TSMOM & \textbf{0.02} & 0.07 & \textbf{0.25} & \textbf{0.23} & \textbf{0.35} & 0.20 & \textbf{0.08} & 0.03 \\
LSTM & -0.01 & 0.05 & -0.24 & -0.32 & -0.33 & 0.28 & -0.04 & 0.07 \\
transformer & 0.00 & \textbf{0.03} & 0.11 & 0.08 & 0.16 & \textbf{0.09} & 0.03 & 0.02 \\
\midrule
\multicolumn{9}{@{}l}{\textit{Metals}}\\
\midrule
$1/N$ & 0.05 & 0.19 & 0.29 & 0.29 & 0.39 & 0.67 & 0.08 & \textbf{0.00} \\
Risk parity & \textbf{0.06} & 0.17 & 0.34 & 0.34 & 0.46 & 0.65 & 0.09 & 0.00 \\
TSMOM & 0.03 & 0.17 & 0.18 & 0.18 & 0.24 & 0.48 & 0.07 & 0.03 \\
LSTM & 0.03 & \textbf{0.11} & 0.30 & 0.26 & 0.39 & \textbf{0.48} & 0.07 & 0.09 \\
transformer & 0.05 & 0.16 & \textbf{0.35} & \textbf{0.35} & \textbf{0.47} & 0.58 & \textbf{0.10} & 0.01 \\
\midrule
\multicolumn{9}{@{}l}{\textit{Energy}}\\
\midrule
$1/N$ & \textbf{0.11} & 0.36 & \textbf{0.30} & \textbf{0.30} & \textbf{0.43} & 1.07 & \textbf{0.10} & \textbf{0.00} \\
Risk parity & 0.09 & 0.33 & 0.28 & 0.28 & 0.40 & 1.06 & 0.09 & 0.00 \\
TSMOM & 0.03 & 0.36 & 0.08 & 0.08 & 0.12 & 1.62 & 0.02 & 0.03 \\
LSTM & 0.06 & \textbf{0.30} & 0.19 & 0.18 & 0.26 & \textbf{0.86} & 0.07 & 0.10 \\
transformer & 0.08 & 0.34 & 0.25 & 0.25 & 0.31 & 1.39 & 0.06 & 0.01 \\
\midrule
\multicolumn{9}{@{}l}{\textit{Agriculture}}\\
\midrule
$1/N$ & 0.03 & 0.26 & 0.12 & 0.12 & 0.18 & 0.83 & 0.04 & \textbf{0.00} \\
Risk parity & 0.04 & 0.26 & 0.16 & 0.16 & 0.25 & 0.76 & 0.05 & 0.00 \\
TSMOM & -0.01 & 0.25 & -0.04 & -0.05 & -0.06 & 1.35 & -0.01 & 0.03 \\
LSTM & 0.01 & \textbf{0.16} & 0.08 & 0.05 & 0.10 & 0.57 & 0.02 & 0.09 \\
transformer & \textbf{0.06} & 0.18 & \textbf{0.34} & \textbf{0.33} & \textbf{0.51} & \textbf{0.48} & \textbf{0.13} & 0.03 \\
\bottomrule
\end{tabular}
\end{table*}

\begin{figure*}[t]
\centering
\includegraphics[width=0.76\textwidth]{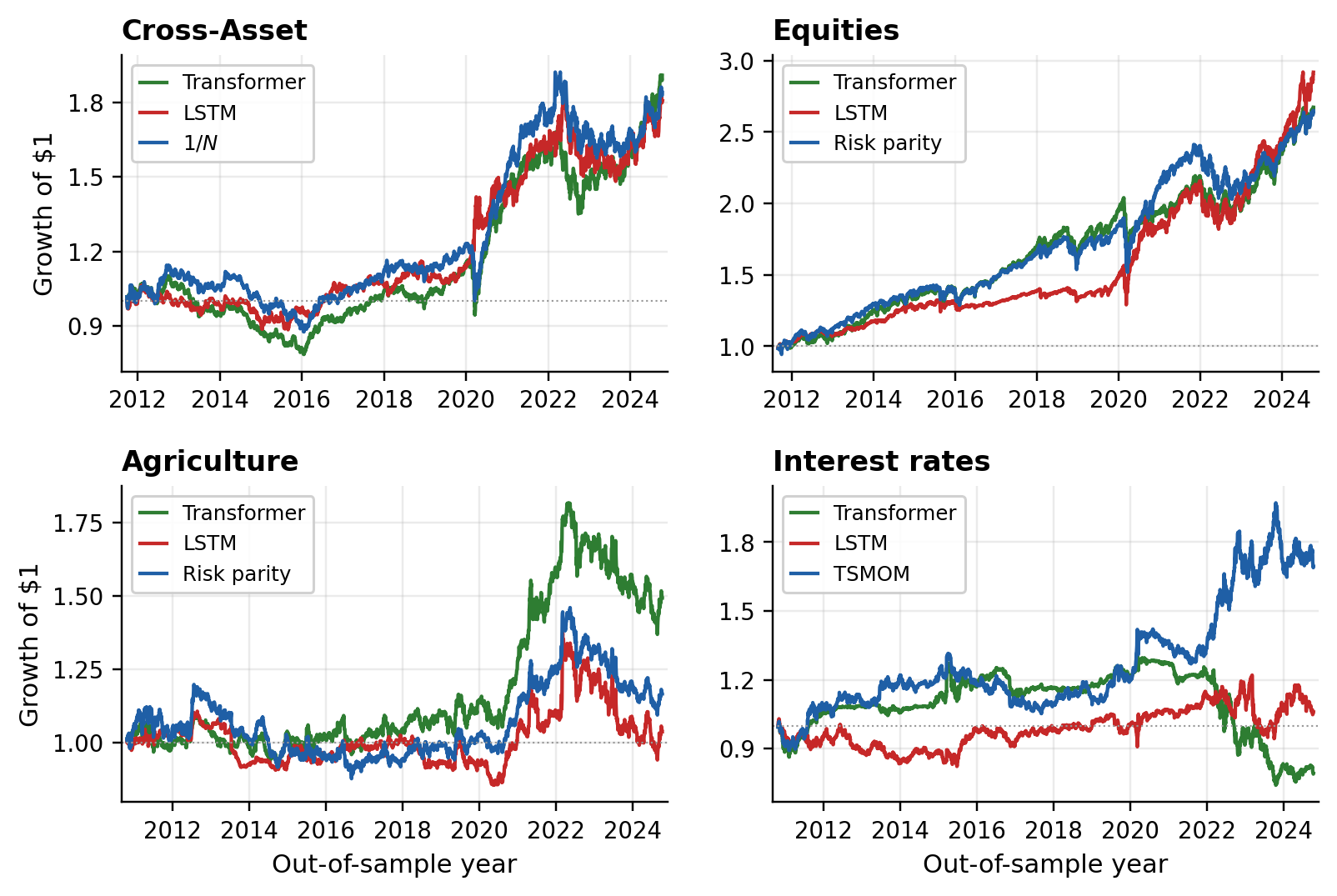}
\caption{Volatility-matched cumulative returns (growth of \$1). Each series is scaled to $10\%$ annual volatility. The learned models are competitive in cross-asset, equities, and agriculture, while time-series momentum leads in interest rates.}
\label{fig:compound}
\end{figure*}

Figure~\ref{fig:compound} plots cumulative returns after scaling each strategy to the same $10\%$ annualized volatility. This normalization is natural for global-macro and managed-futures portfolios, where futures exposures can be scaled through notional and managers often allocate risk rather than dollars across sleeves. It puts every strategy on the same risk scale, so the figure compares compounding per unit of risk. The learned models track or exceed the best rule over most of the sample in the cross-asset, equity-index, and agriculture universes. They trail most clearly in interest rates, where time-series momentum performs best.

\enlargethispage{-0.7in}
\subsection{Risk-Adjusted Performance and Transaction Costs}

\begin{figure}[H]
\centering
\includegraphics[width=0.93\columnwidth]{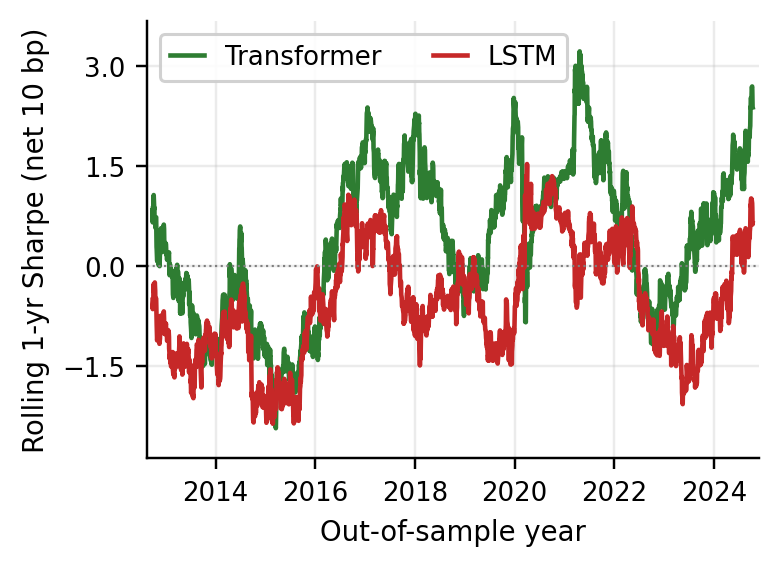}
\caption{Rolling one-year ($252$-day) Sharpe ratio on the cross-asset portfolio, net of a $10$~bp transaction cost. The low-turnover transformer is less eroded by trading costs than the LSTM.}
\label{fig:rolling}
\end{figure}

The two end-to-end models differ sharply in turnover, measured as the average daily one-sided change in target weights, $\tfrac12\sum_i |w_{i,t}-w_{i,t-1}|$. The transformer holds steadier positions, with turnover near $0.02$ per day across universes. The LSTM trades much more, with turnover between $0.07$ and $0.17$. On the cross-asset portfolio, turnover is $0.17$ for the LSTM versus $0.02$ for the transformer. Figure~\ref{fig:rolling} shows the effect of this turnover difference. After transaction costs, the transformer's rolling one-year Sharpe stays above the LSTM's over most of the sample.

The cross-asset results make the cost effect clear. Before costs, the transformer has a slightly higher Sharpe than the LSTM, $0.55$ versus $0.50$. After applying the $2$~bp transaction-cost rate used for liquid futures as in~\cite{kisiel2023portfoliotransformers,cucuringu2021universalendtoendapproachportfolio}, the LSTM's Sharpe falls from $0.50$ to $0.33$. The transformer's Sharpe falls only from $0.55$ to $0.54$, which is the highest net Sharpe of any strategy in Table~\ref{tab:cost}. The gap widens as costs increase. At $10$~bp, the LSTM falls to $-0.38$ on cross-asset, while the transformer remains near $0.50$. The ranking between the two models is stable across the $1$--$10$~bp cost range.

\begin{table}[H]
\centering
\caption{Net Sharpe ratio of the cross-asset portfolio as transaction costs rise. The transformer is much less sensitive to trading costs because of its low turnover, whereas the LSTM deteriorates quickly.}
\label{tab:cost}
\small
\begin{tabular}{@{}lrrrrr@{}}
\toprule
& \multicolumn{5}{c}{Transaction cost (bp)} \\
\cmidrule(lr){2-6}
Strategy & $0$ & $1$ & $2$ & $5$ & $10$ \\
\midrule
$1/N$ & 0.52 & 0.52 & 0.52 & \textbf{0.52} & \textbf{0.52} \\
Risk parity & 0.15 & 0.14 & 0.14 & 0.13 & 0.11 \\
TSMOM & 0.37 & 0.36 & 0.35 & 0.33 & 0.29 \\
\addlinespace
LSTM & 0.50 & 0.42 & 0.33 & 0.06 & -0.38 \\
transformer & \textbf{0.55} & \textbf{0.54} & \textbf{0.54} & \textbf{0.52} & 0.50 \\
\bottomrule
\end{tabular}
\end{table}

The LSTM is still competitive in some settings. It is strongest in the equity index, where it records the highest Sharpe in the study at $0.87$. Across most universes, however, the transformer is the better model to run. It has stronger gross performance in several sleeves and is much cheaper to trade. These net results charge transaction costs at evaluation on policies trained gross of costs. We also tested cost-aware training by adding an explicit turnover penalty to the loss function. This reduced LSTM trading, but did not improve net performance enough to change the ranking. For the already low-turnover transformer, the penalty was too restrictive. The gross-trained transformer therefore remains the model we would run.

\subsection{Skill versus Market Exposure}

The equity-index results are the only learned-policy Sharpe ratios that remain clearly different from zero after the Bonferroni correction. We therefore ask whether those results reflect residual timing skill, market exposure, or a more efficient form of benchmark exposure. To do this, we regress each model's return on the diversified benchmark for the same universe and decompose performance into a market-exposure component, $\beta$, and a residual alpha (Table~\ref{tab:alphabeta}).

For the LSTM, the equity-index alpha remains significant using Newey-West standard errors. For the transformer, it does not. The transformer's equity alpha is small and statistically insignificant, while its market-exposure component explains most of the return. No other universe shows a residual alpha that is distinguishable from zero. Direct tests for market-timing ability are also insignificant.

The transformer's equity result is therefore not best interpreted as market-neutral alpha. It is closer to a risk-efficient directional exposure as the model earns a Sharpe ratio comparable to the equity benchmark while carrying a sub-unit benchmark beta. This is still useful from a portfolio-construction perspective, but it is different from residual alpha. The two architectures also differ in their attribution. The LSTM has a larger residual alpha in the equity index, while the transformer has larger residual alpha estimates in metals and energy. Skill-versus-exposure attribution is therefore model-specific. It must be measured on the architecture in use and should not be assumed to transfer from one model to another.

\begin{table}[H]
\centering
\caption{Per-universe $\alpha$/$\beta$ decomposition against each universe's $1/N$ benchmark. $\alpha$ is annualized in percent from a Newey--West HAC regression, and \textsuperscript{*} indicates significance at $5\%$.}
\label{tab:alphabeta}
\small
\begin{tabular}{@{}lrrrr@{}}
\toprule
 & \multicolumn{2}{c}{LSTM} & \multicolumn{2}{c}{transformer} \\
\cmidrule(lr){2-3}\cmidrule(lr){4-5}
Universe & $\alpha$ & $\beta$ & $\alpha$ & $\beta$ \\
\midrule
Equity & $+3.9$\textsuperscript{*} & $0.50$\textsuperscript{*} & $+0.8$ & $0.63$\textsuperscript{*} \\
Metals & $-0.1$ & $0.41$\textsuperscript{*} & $+1.4$ & $0.67$\textsuperscript{*} \\
Agri. & $-4.9$ & $0.38$\textsuperscript{*} & $+2.0$ & $0.47$\textsuperscript{*} \\
FX & $-0.6$ & $0.31$\textsuperscript{*} & $+0.5$ & $0.22$\textsuperscript{*} \\
Rates & $+0.1$ & $0.18$\textsuperscript{*} & $-0.3$ & $0.52$\textsuperscript{*} \\
Energy & $-0.4$ & $0.56$\textsuperscript{*} & $+0.3$ & $0.75$\textsuperscript{*} \\
\bottomrule
\end{tabular}
\end{table}

\subsection{Does Added Complexity Help?}

The pooled cross-asset transformer is sensitive to random initialization. A single seed can produce a Sharpe ratio that varies widely, and its performance is not reliably distinguishable from zero. To reduce this source of noise, we average three independently seeded transformer models. This removes much of the seed-level variance and makes the allocator competitive with the standard rules. We also tested larger seed ensembles, but adding more than three seeds did not materially change the results.

Seed averaging is the only enhancement that reliably helps in our tests. The differences across universes in Table~\ref{tab:results} suggest several natural extensions, including a mixture-of-experts model, a larger feature set, per-class tuning, and an ensemble across the LSTM and transformer. We tested these ideas in relatively simple, out-of-the-box implementations rather than as fully optimized model classes. These tests did not produce materially different out-of-sample results after transaction costs.

The cross-model ensemble illustrates the pattern. The transformer and LSTM returns are positively correlated, and both models are noisy. As a result, an equal-weighted ensemble tends to fall between the two models rather than outperforming both. We also tested a simple rule that chooses the better model over time, but this added estimation noise and did not improve performance. The turnover-penalized loss produced a similar result. It reduced trading for the high-turnover LSTM, but it over-constrained the already low-turnover transformer. Overall, the evidence favors the simple seed-averaged transformer over the additional complexity we tested.

\subsection{Why Cross-Asset Timing Is Hard}

We read the modest performance as confirmation that the problem is indeed quite challenging, not that the model fails to optimize. The in-sample study in Section~\ref{sec:insample} shows that the objective improves and a train-validation gap opens. The loss function and weight layer therefore appear to work as intended. The harder question is why the best out-of-sample results are still modest.

Several features of the setting limit what the models can achieve. First, the liquid-universe design removes illiquidity as a likely source of return. That gives a cleaner test of timing skill, but it also leaves a thinner signal to trade. Second, the Sharpe objective has a natural solution in markets that drift upward over the sample. A model can improve the objective by learning a steady, sub-unit exposure to risk, even if it finds little residual alpha. In this sense, some results look closer to risk-efficient beta exposure than to market-neutral timing skill. Third, the single-asset-class sleeves are small. They contain only two to four contracts, which leaves little cross-section for the models to exploit. The sixteen-contract cross-asset universe is the widest universe we form, and it is where the transformer most clearly benefits from breadth, although the highest absolute learned-model Sharpe occurs in the equity-index sleeve. Fourth, the benchmark is demanding. Equal weighting and risk parity are hard to beat on liquid futures, so a learned policy has to clear a high practical bar before it shows value.

These same forces explain why no strategy wins everywhere. The best allocator depends on the asset class and the regime. Time-series momentum is hardest to beat in rates and currencies. Naive diversification performs well in the trending equity and energy sleeves. The learned policy performs best in the broad cross-asset portfolio and in the choppier commodity classes. This dependence is central to global-macro and managed-futures allocation, where the right model can differ across sleeves and where the tradeoff between timing and diversification changes over time. It also limits what we can claim. We study one daily frequency on a highly liquid universe. At higher frequency, or across a wider set of contracts, an end-to-end policy may have more signal and more breadth to exploit. The daily liquid setting is therefore best viewed as a conservative test of the framework rather than a setting designed to maximize its advantage.

\section{Conclusion and Future Work}
\label{sec:conclusion}

We treat this study as a conservative test of end-to-end allocation in a liquid cross-asset setting applicable to global-macro or managed-futures systematic strategies. The evidence is useful, but not uniform. Across the seven universes and the configurations we tried, the transformer produces one learned-policy result that is both statistically significant after correction and stable across seeds: the equity-index sleeve. Even there, the decomposition shows that most of the performance comes from market exposure rather than residual timing skill.

This still leaves a useful allocator. On the full sixteen-contract cross-asset portfolio, the transformer attains the highest Sharpe ratio of any strategy, slightly above equal weighting. It also trades much less than the LSTM, so its performance is largely preserved after realistic transaction costs. The result is not standalone, market-neutral alpha. It is closer to a low-turnover, risk-efficient allocation rule that can compete with simple benchmarks in some settings.

Two lessons extend beyond this study. First, learned allocation policies should be benchmarked against the rules a practitioner would actually run, not against $1/N$ alone. Second, risk, turnover, and attribution should be measured on the exact model in use. The LSTM and transformer differ not only in performance, but also in turnover and in the split between beta exposure and residual alpha. In practice, this means evaluating costs by asset class and averaging across seeds before committing capital.

Several limitations apply. For the transformer, turnover is low, around $0.01$--$0.03$ per day, so gross and net performance are close. The results are still seed-sensitive, so individual estimates should be treated as indicative rather than definitive. We tuned the training recipe mostly on the faster LSTM, so transformer-specific tuning could change the attribution. We also study one daily frequency and two architectures, not a broader model class.

The natural extensions follow directly from these limitations. Higher-frequency data may give the policy more signal to learn from. A larger futures universe, with more contracts per asset class and more asset classes, would give the model more breadth to exploit. Contract and asset-class embeddings could then allow one model to share information across related instruments while still learning contract-specific behavior.


\end{document}